\documentclass[APA,Times1COL]{WileyNJD-v5} 

\articletype{Original Article}%

\startpage{1}

\usepackage{natbib}
\usepackage{amsthm}
\usepackage{multirow}
\usepackage{amsmath}
\usepackage{enumitem}
\usepackage{arydshln}

\usepackage[nomarkers, nolists]{endfloat}

\newcommand{\bB}{\mathbf{B}}

\newcommand{\bW}{\mathbf{W}}

\newcommand{\sB}{\mathcal{B}}

\newcommand{\sD}{\mathcal{D}}

\newcommand{\sL}{\mathcal{L}}

\newcommand{\sP}{\mathcal{P}}

\newcommand{\sR}{\mathcal{R}}
\newcommand{\sS}{\mathcal{S}}

\newcommand{\bbP}{\mathbb{P}}

\newcommand{\bbR}{\mathbb{R}}

\newcommand{\E}{\mathbb{E}}

\newcommand{\tp}{\text{T}}

\newcommand{\btheta}{\boldsymbol{\theta}}

\begin{document}

\title{Hazard and Beyond: Exploring Five Distributional Representations of Accelerometry Data for Disability Discrimination in Multiple Sclerosis
}

\author[1]{Pratim Guha Niyogi*}
\author[2]{Muraleetharan Sanjayan}
\author[3]{Dmitri Volfson}
\author[2,4]{Kathryn C. Fitzgerald}
\author[2,4]{Ellen M. Mowry}
\author[1]{Vadim Zipunnikov}

\authormark{P. Guha Niyogi \textsc{et al}}

\address[1]{\orgdiv{Department of Biostatistics}, \orgname{Johns Hopkins University}, \orgaddress{\state{Baltimore, MD}, \country{USA}}}

\address[2]{\orgdiv{Department of Neurology}, \orgname{Johns Hopkins University}, \orgaddress{\state{Baltimore, MD}, \country{USA}}}

\address[3]{\orgdiv{Statistical and Quantitative Sciences}, \orgname{Takeda
Development Center Americas, Inc}, \orgaddress{\state{Cambridge, MA}, \country{USA}}}

\address[4]{\orgdiv{Department of Epidemiology}, \orgname{Johns Hopkins University}, \orgaddress{\state{Baltimore, MD}, \country{USA}}}

\corres{*Pratim Guha Niyogi, \email{pnyogi1@jhmi.edu}}

\presentaddress{615 North Wolfe Street,
Baltimore, MD, 21205.}

\abstract[Abstract]{
Research on modeling the distributional aspects in sensor-based digital health (sDHT) data has grown significantly in recent years. Most existing approaches focus on using individual-specific density or quantile functions. However, there has been limited exploration to assess the practical utility of alternative distributional representations in clinical contexts collecting sDHT data.
\par
This study is motivated by accelerometry data collected on 246 individuals with multiple sclerosis (MS) representing a wide range of disability (Expanded Disability Status Scale, EDSS: 0-7). We consider five different individual-level distributional representations of minute-level activity counts: density, survival, hazard, quantile, and total time on test functions. For each of the five distributional representations, scalar-on-function regression fits linear discriminators for binary and continuously measured MS disability, and cross-validated discriminatory performance of these linear discriminators is compared across. 
\par
The results show that individual-level hazard functions provide the highest discriminatory accuracy, more than double the accuracy compared to density functions. Individual-level quantile functions provided the second-highest discriminatory accuracy. These findings highlight the importance of focusing on distributional representations that capture the tail behavior of distributions when analyzing digital health data, especially in clinical contexts.
}


\maketitle

\footnotetext{\textbf{Abbreviations:} MS, Multiple Sclerosis; PA, physical activity; DF, Distribution function; QF, quantile function; TTT, total time on test; ROC, Receiver Operating Characteristic; AUC, Area Under the Curve 
}

\section{Introduction}
\label{sec:intro}

Sensor-based digital health technologies (sDHTs) continuously collect vast amounts of rich and complex data on individual behaviors and health. sDHT data including continuously-monitored physical activity, heart rate, glucose levels, is often heterogeneous across individuals. Traditional modeling approaches methods summarize sDHTs data into simple scalar measures such as total activity count, average heart rate, or total daily steps. To better capture the full individual-specific variability provided by digital health technologies, ``distributional data analysis'' became a popular alternative \citep{petersen2019frechet, ghosal2021distributional}.

In distributional data modeling, instead of reducing data to summary statistics, each individual’s data is represented as a distribution, capturing the variability of their epoch-level sDHT data over the assessment period. Thus, individual-level distributions of sDHT date can serve as outcomes, predictors, or both in regression frameworks known as ``distribution regression'' \citep{oliva2013distribution, petersen2016functional, szabo2016learning, fang2020optimal, chen2021wasserstein}.

In many clinical studies of multiple sclerosis (MS)—a chronic neurological disease affecting the central nervous system —scientists often study the association between disability and ambulatory physical activity assessed via wearable accelerometry \citep{ezeugwu2015mobility, abbadessa2021assessment, keller2022using}. The most commonly used summary measures for physical activity include total log activity counts, minutes of moderate-to-vigorous-intensity physical activity, and active-to-moderate sedentary transition probabilities \citep{leroux2019organizing, smirnova2020predictive}. As discussed above, these scalar measures ignore the distributional aspect of accelerometry data.

The analysis of distributional data typically relies on classical functional analysis tools, with density functions representing distributions. However, densities are rarely fully observed. Almost always there is a random sample of data, and the density must be estimated using common methods such as kernel density estimation \citep{poczos2013distribution, oliva2013distribution}. Some functional compositional methods for analyzing samples of densities were developed by \citet{petersen2016functional} and \citet{hron2016simplicial}. It is well known that due to the boundary effect \citep{petersen2016functional}, kernel density estimation methods often perform inadequately. Later, \citet{augustin2017modelling} proposed methods that summarized individual-level distributions via histograms. These distributional representations were used as covariates in linear regression models. Intuitively, discretizing histograms and estimating additional tuning parameters may introduce discrepancies in parameter selection steps.

\citet{chen2021wasserstein} and \citet{ghodrati2022distribution} proposed distribution-on-distribution regression models utilizing Wasserstein distances and optimal transport ideas. \citet{talska2021compositional} developed a compositional scalar-on-function regression approach using a centered log-ratio transformation of subject-specific densities. This involves mapping densities from the Bayes space of density functions to a Hilbert $L_{2}$ space and applying traditional functional data modeling. \citet{matabuena2021glucodensities} suggested using subject-specific densities of glucose measures, obtained through continuous glucose monitoring, as predictors and responses. They demonstrated the advantages of this approach over the use of summary measures typically employed in continuous glucose monitoring studies by developing a non-parametric kernel functional regression model using a 2-Wasserstein distance to model scalar outcomes and glucose density predictors.

\citet{yang2020quantile} proposed quantile-function-on-scalar regression approaches for modeling quantile functions as outcomes, though this approach imposes constraints requiring the regression approximant to be a valid quantile function. Recently, \citet{ghosal2021distributional} proposed a scalar-on-functional regression framework using subject-specific quantile functions as predictors, while also employing a generalized additive model to capture possible non-linear effects of quantile functions. In contrast to nonlinear space analyses of density functions \citep{talska2021compositional} and nonparametric methods using distances or kernels \citep{matabuena2021distributional}, the approach of \citet{ghosal2021distributional} is semi-parametric, offering direct interpretability based on subject-specific distributions at quantile levels. Recently, \citet{petersen2022modeling} provided a comprehensive review of the latest developments in this area in a tutorial-style format.

In reliability and survival analysis, non-negative random variables arise in a wide variety of applications. Mathematical functions such as the distribution function, survival function, density, hazard rates, and total time on test are useful for characterizing probabilistic properties, each with different interpretations (see \citet{marshall2007life, nair2013quantile} for more details). Although there have been isolated instances where the replacement of distribution functions with the other aforementioned functions has been attempted, systematic comparison has not been done. 
The main contribution of this article is comparing five distributional representations of minute-level activity counts and demonstrating significantly stronger clinical utility of hazard function, the distributional representation that has received relatively little attention in literature.
 
 The article is organized as follows. In Section \ref{sec:motivating-example}, we discuss the motivation study of multiple sclerosis. In Section \ref{sec:model}, we discuss five different distributional representations and illustrate the proposed methodology. Section \ref{sec:data-analysis} provides a detailed statistical analysis. Finally, in Section \ref{sec:dis}, we conclude with a discussion of the findings, limitations, and potential future directions.

\section{HEAL-MS study}
\label{sec:motivating-example}

In this section, we present a motivating example and discuss a study conducted within the Johns Hopkins MS Precision Medicine Center of Excellence, which aimed to better understand the manifestation of disability in multiple sclerosis (MS) represented in ambulatory physical activity assessed via wearable accelerometers.

The study was conducted at Johns Hopkins University from January 2021 to March 2023 as part of the Home-based Evaluation of Actigraphy to Predict Longitudinal Function in Multiple Sclerosis (HEAL-MS) project \citep{rjeily2022using, tian2020longitudinal}. MS is a complex neurological disorder affecting the brain and spinal cord, with diverse symptoms that can severely impact quality of life. In early MS, the myelin sheath around axons in the central nervous system is attacked, while later in the course, there is degeneration of the axons themselves. It is this latter process that is thought to underlie the slowly worsening disability that affects many with MS, known as ``progressive'' MS. 
The Expanded Disability Status Scale (EDSS) is a standard clinical tool for measuring MS disability, ranging from 0 (normal neurological function) to 10. Scores from 1.0 to 3.5 indicate good mobility (``low EDSS''), while scores of 4 or higher signify loss of mobility (``high EDSS'').

The study's goal was to explore the relationship between EDSS scores and ambulatory physical activity. Participants wore accelerometers (GT9X Actigraph) on their non-dominant wrist for two weeks.  Raw data were collected at a 30Hz sampling rate, processed using ActiLife v6.134 Lite, and summarized via minute-level activity counts. For the purpose of this study, we focused on the 12 hours of typical wake time that was defined as the period between 8am to 8pm. To account for skewness, we compare both the original activity counts and their log-transformed version defined by $x \rightarrow \log(x+1)$ transformation.

The analysis included 246 participants, all aged at least 40 (mean age 54.8 years, SD 8.6, 70.7\% are female), with no known comorbidities affecting physical activity. Of the sample, 158 participants (64\%) with an average age of 53.8 (8.3) had low EDSS scores, and 88 (36\%) had high EDSS scores with average age of 56.8 (8.7). Moreover, 113 (71.5\%) and 61 (69.3\%) are female for the groups low and high EDSS respectively. 

\section{Modeling frameworks}
\label{sec:model}

In the following subsections, we discuss five distributional representations, how each of them are estimated from sDHT data and how those distributional representations are used in scalar-on-distribution functional regression. 
\subsection{Five distributional representations}
\label{subsec:dist}
Let random density is a measurable map $f: (\Omega, \sS, P) \rightarrow (\sD, \sB_{\sD})$ where $(\Omega, \sS, P)$ is a probability space and $\sB_{\sD}$ contains the Borel sets on $(\sD, d)$. We assume that a sample of independent and identically distributed random densities $f_{i}$ is available for regression analysis, but in reality, we observe them rarely. 
Instead, we observe the collection of scalars $X_{i1}, \cdots, X_{im_{i}}$ for $i = 1, \cdots, n$ where $\{ X_{ij}: j = 1, \cdots, m_{i}\} \sim f_{i}$. Throughout this paper, we assume that $X_{ij}$ are non-negative continuous random variables and this assumption is very trivial in our data example. 
For subject $i$, the function $S_{i}$ is defined as $1-F_{i}(x) = \bbP\{ X_{i} > x\}$ and is termed the survival function. The survival function represents the probability that a specific event has not occurred by a certain time point $x$. 
Here $F_{i}$ indicates the subject-specific distribution function (DF). 
Hazard rate (HR) of $X_{i}$ is defined as $\lambda_{i}(x) = \lim_{\delta\downarrow 0}\frac{\bbP\{x \leq X_{i} < x + \delta| X_{i} > x\}}{\delta}$ which is the conditional probability that an event will fail the next small interval of time given that the event has been survived at $x$. If $F_{i}$ is an absolutely continuous DF with density $f_{i}$, the HR for subject $i$ is defined as $(-\infty, \infty)$ by $\lambda_{i}(x) = f_{i}(x)/S_{i}(x)\textbf{1}\{S_{i}(x) > 0\} + \infty\textbf{1}\{S_{i}(x) = 0\}$ where $\textbf{1}(x \in A)$ takes value $1$ when $x \in A$ and zero otherwise. Moreover, we define the subject-specific quantile function (QF) $Q_{i}(p) = \inf \{ x : F_{i}(x) \geq p\}$ for $p \in [0, 1]$. In this article, we restrict our attention to the cases where both $F_{i}(x)$ and $Q_{i}(p)$ are continuous, which ensures that $Q_{i} = F_{i}^{-1}$ satisfying $F_{i}(Q_{i}(p)) = p$ and $Q_{i}(F_{i}(x)) = x$, which confirms that the QF and DF exhibit strictly ascending behavior within their individual domains.
\par
In addition to the density, survival, hazard, and quantile functions, we also can use the total time on test (TTT) \citep{epstein1953life, brunk1972statistical} function as a choice of the distributional representation. This transformation based on the quantiles was developed in the early 1970s in reliability problems. It has since been applied in other areas such as stochastic modeling, maintenance scheduling, risk assessment, and energy sales. The subject-specific TTT transform of the distribution function $F$, viz., $T_{i}(p)$ is defined on the interval $[0, 1]$ as $T_{i}(p) = \int_{0}^{Q_{i}(p)}\left\{ 1- F_{i}(x) \right\}dx$. For convenience, we list some of the basic properties below. 
\begin{itemize}
    \item For each $i$, like quantile function, $T_{i}$ is defined on $[0, 1]$ which is independent of the subject. 
    By definition, $T_{i}$ starts from zero, and more importantly, if the mean of the distribution exits, then $Q(1) = \infty$, and as a result, $T_{i}(1)$ is the mean of the underlying random variable. 
    \item Since we have assumed that the DF is continuous, TTT is an increasing function (moreover the converse is also true).
    \item TTT is a quantile function and the corresponding distribution is termed as the transformed distribution, $T_{i}(p) = \psi_{i}(Q_{i}(p))$ where $\psi_{i}(x) = \int_{0}^{x}S_{i}(x)dx$.
    \item Moreover the baseline DF $F_{i}$ and QF can uniquely determined by $T_{i}(p)$.
\end{itemize}
\par
However, in principle, $\E\{ X_{i} \}$ is a scalar quantity that can be obtained from integrating survival or quantile function over its appropriate range. Therefore, instead of using such a naive summary measure, we would like to consider a function defined on a closed and bounded interval on $\bbR$ with the terminal point coinciding with the expectation of the underlying random variable (we assume throughout the expectation exits).

\begin{remark}[Distributional morphology]
The QF can completely characterize the distribution of individual observation. Moreover, the DF determines the TTT and the reverse is also true. In addition, if $F_{i}$ has hazard rate $\lambda_{i}$ then $T_{i}$ is differentiable and $\frac{\partial T_{i}(p)}{dp}|_{p = F(x)} = \frac{1}{\lambda_{i}(p)}$. For example, we consider exponentiated Weibull distribution parameterized by scale parameter $\lambda_{0}$, shape parameter $\kappa_{0}$ and power parameter $\alpha_{0}$ with density function 
\begin{equation}
    f(x; \lambda_{0}, \kappa_{0}, \alpha_{0}) = \frac{\alpha_{0}\kappa_{0}}{\lambda_{0}}
    \left\{\frac{x}{\lambda_{0}} \right\}^{\kappa_{0}-1}
    \left\{1-\exp\{-(x/\lambda_{0})^{\kappa_{0}}\} \right\}^{\alpha_{0}-1}
    \exp\{-(x/\lambda_{0})^{\kappa_{0}}\}\textbf{1}\{x \geq 0\},
\end{equation}
where $\lambda_{0}, \kappa_{0}, \alpha_{0} > 0$ and $\textbf{1}\{x\in A\}$ is denoted as indicator function that takes value 1 if $x\in A$ and 0 otherwise. We consider the following choices of parameters to demonstrate different shapes of the hazard function. (i) for constant HR $(\lambda_{0}, \kappa_{0}, \alpha_{0})  = (20, 1, 1)$, (ii) for decreasing HR, $(\lambda_{0}, \kappa_{0}, \alpha_{0}) = (5,0.5,2)$, (iii) for increasing HR $(\lambda_{0}, \kappa_{0}, \alpha_{0}) = (50,2,2)$, (iv) for bathtub-shaped HR $(\lambda_{0}, \kappa_{0}, \alpha_{0}) =(150,4,0.15)$ and (v) for unimodal HR $(\lambda_{0}, \kappa_{0}, \alpha_{0}) = (5, 0.55, 4)$. Figure \ref{fig:dist} demonstrates the different distributional representations for these parametric families.

To illustrate these five distributional representations for accelerometry data from individuals with MS, Figures 2 (original activity counts) and 3 (log-transformed activity counts) show five distributional representations for two representative subjects:  one with a low EDSS score of 1 (minimal disability, shown in blue) and another with a high EDSS score of 6 (greater disability, shown in red). The low EDSS subject exhibits broader distributions, slower declines in survival, and more sustained activity at higher percentiles, indicating greater variability and higher intensity in physical activity. In contrast, the high EDSS subject’s curves are more "compressed", with sharper survival declines and hazard peaks at lower activity levels.

\begin{figure}
    \centering
    \includegraphics[width = \textwidth]{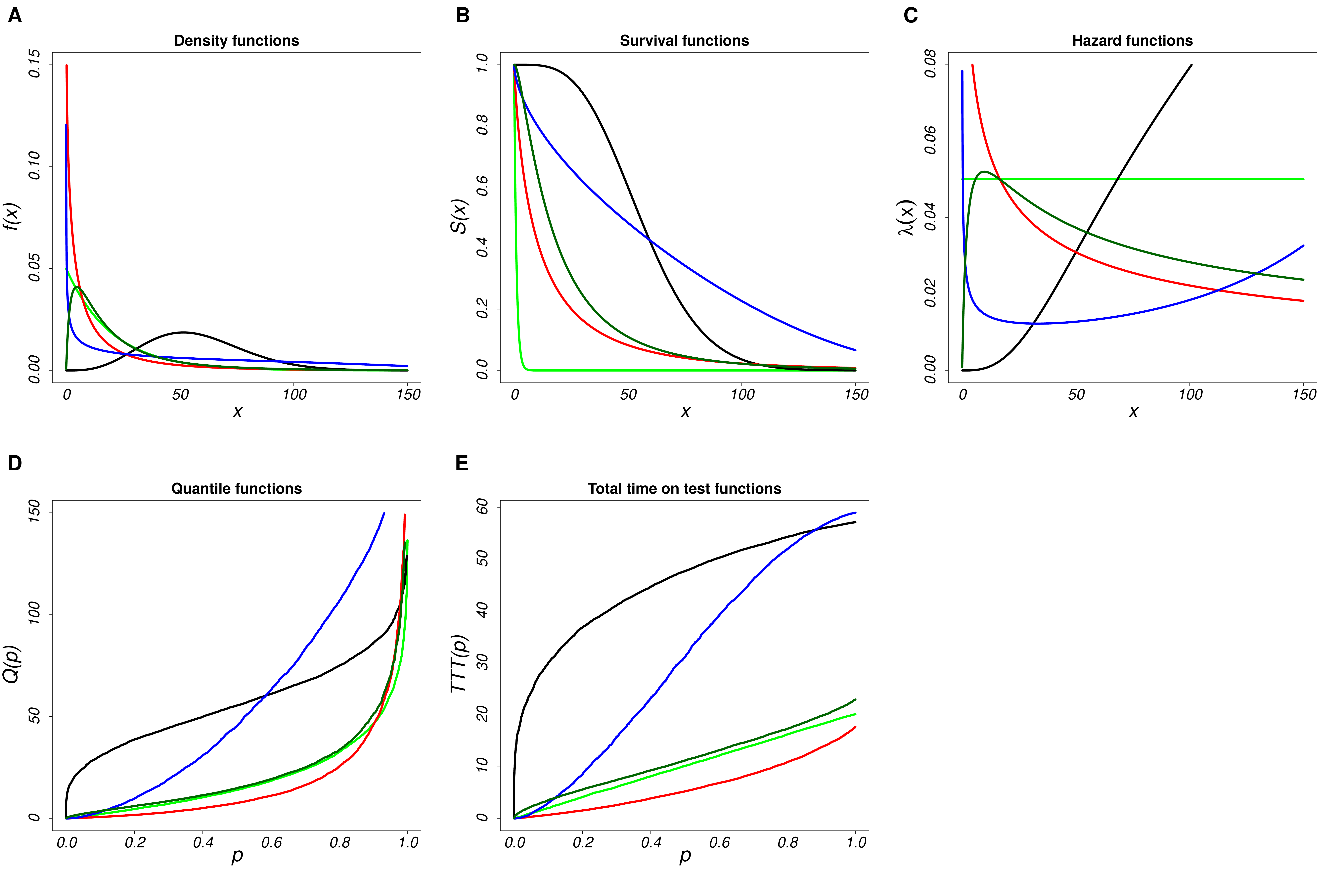}
    \caption{\underline{\textbf{Distributional morphology:}} Density functions (A), survival functions (B), hazard functions (C) with corresponding quantile functions (D), and corresponding TTT functions (E). Here the underlying distribution is an exponentiated Weibull distribution and we consider the different scale, shape, and power parameters so that we obtain the different shapes of the hazard functions.}
    \label{fig:dist}
\end{figure}
\end{remark}

\subsection{Estimation of five distributional representations}
To pursue statistical modeling with the distributional representations mentioned in Section \ref{subsec:dist}, we need to estimate those statistical objects. Here we summarize the estimation procedures of the distribution representations. The density function is estimated by the kernel density estimator as described below. 
\begin{equation}
    \widehat{f}_{i, h_{i}}(x) = \frac{1}{h_{i}}\sum_{j=1}^{m_{i}}K\left(\frac{x - X_{ij}}{h_{i}} \right),
\end{equation}
where $K$ is the non-negative kernel function and $h_{i} > 0$ is a smoothing parameter (a.k.a bandwidth). Under the choice of the Gaussian kernel, the rule of thumb of selecting the subject-specific bandwidth is $h_{i} \sim 1.06\widehat{\sigma}_{i}m_{i}^{-1/5}$ where the standard deviation is estimated by $X_{i1}, \cdots, X_{im_{i}}$ \citep{silverman2018density}. The Kaplan-Meier estimator is used for estimating survival function \cite{kaplan1958nonparametric}, given by 
\begin{equation}
 \widehat{S}_{i}(x) = \prod_{j: X_{ij} \leq x}\left(1-\frac{D_{ij}}{N_{ij}} \right),   
\end{equation}
where for each subject $i$, $N_{ij}$ is the number of events occurred after $X_{ij}$ and $D_{ij}$ is the number of events occurred at the interval $[X_{i,j-1}, X_{ij})$. Another powerful estimator is Nelson-Aalen estimator \citep{nelson1972theory, aalen1978nonparametric} which can produce an estimator for cumulative hazard rate function $\Lambda(x) = \int_{0}^{x}\lambda(t)dt = -\log\{S(x)\}$, where $\lambda_{i}(x) = \lim_{\delta\downarrow 0}\frac{\bbP\{x \leq X_{i} < x + \delta| X_{i} > x\}}{\delta}$. It can be shown that $\widehat{\Lambda}_{i} = \prod_{j: X_{ij} \leq x}\frac{D_{ij}}{N_{ij}}$. At the same time, due to the definition of the integral, $\Lambda_{i}(x) \approx \sum_{\text{grid points $t$ such that $t+\delta$}} \lambda_{i}(t)\delta$, by definition of the hazard function, $\lambda_{i}(x)\delta \approx \bbP\{ x \leq X < x + \delta| X > \delta\}$ could be estimated by $D_{ij}/N_{ij}$, reasonably. Quantile functions can be estimated by the linear interpolation of the order statistics, which is not required for bandwidth selection in kernel density estimation \citet{parzen2004quantile}. We use the following estimator of quantile functions at level $p$, 
\begin{equation}
    \widehat{Q}_{i}(p) = (1-\omega)X_{i, [(n+1)p]} + \omega X_{i, [(n+1)p] + 1},
\end{equation}
where the weight $\omega$ satisfies $(n+1)p = [(n+1)p] + \omega$ and $[x]$ denotes the greatest integer that does not greater than $x$. In most statistical software, tools are readily available to generate estimates of the underlying distributional representations. For instance, in R, the functions \texttt{density} and \texttt{quantile} from the \textit{stats} package \citep{stats} are employed to estimate density and quantile functions, while the \texttt{survfit} function from the \textit{survival} package \citep{survival-book} is used to estimate survival and hazard functions.
\par
Recall that, the total time on test (TTT) for each subject represents the total time the subject survives from the beginning of the observation period until the event of interest occurs. 
In other words, in a life-length study involving multiple units, some units will fail while others will survive the test period, and the total time on test statistic is the sum of all observed and incomplete life lengths.
When the number of units tested approaches infinity, the limit of this statistic is known as the total time on test transform \citep{epstein1953life, brunk1972statistical}.
Formally, for subject $i$, suppose there are $m_{i}$ items to test and the successive failures are observed at $X_{i,1:m_{i}} \leq X_{i, 2:m_{i}} \leq \cdots \leq X_{i, m_{i}:m_{i}}$ are the order statistics from a sample $X_{i1}, \cdots \leq X_{im_{i}}$  with absolutely continuous DF $F_{i}$. 
Note that the test time observed between $0$ and $X_{i, 1:m_{i}}$ is $m_{i}X_{i, 1:m_{i}}$, that of between $X_{i, 1:m_{i}}$ and $X_{i, 2:m_{i}}$ is $(m_{i}-1)(X_{i, 2:m_{i}}-X_{i, 1:m_{i}})$, etc.
Therefore, we can calculate the total time on test statistic (TTT-stat) during $(0, x)$, for $x \in (X_{i, r:m_{i}},X_{i, r+1:m_{i}}]$ as 
\begin{equation}
\label{eq:TTTlim}
    \tau_{i}(x) = m_{i}X_{i, 1:m_{i}} + (m_{i}-1)(X_{i, 2:m_{i}}-X_{i, 1:m_{i}}) + \cdots + (m_{i}-r)(x - X_{i, r:m_{i}}).
\end{equation}
It is easy to observe that an estimate of the quantile function $\widehat{Q}_{i}(p) = \inf\{x: \widehat{F}_{i}(x) \geq p\}$, for the empirical distribution function $\widehat{F}_{i}$
\begin{equation}
    \widehat{F}_{i}(x) = 
    \begin{cases}
    0, & x \leq X_{i, 1:m_{i}}\\
    p_{ij} = j/m_{i}, & X_{i, :m_{i}} < x \leq X_{i, (j+1):m_{i}} \text{ for } j = 1, \cdots, m_{i}-1 \\
    1, & x \geq X_{i, m_{i}:m_{i}}\\
    \end{cases}
\end{equation}
we have for fixed $i$,  
\begin{equation}
 \lim_{m_{i} \rightarrow \infty}\lim_{r/m_{i} \rightarrow p}
 \tau_{i}(X_{i, r:m_{i}})/m_{i} = \lim_{m_{i} \rightarrow \infty}\lim_{r/m_{i} \rightarrow p}
 \int_{0}^{\widehat{Q}_{i}(r/m_{i})}\{ 1 - \widehat{F}_{i}(x)\}dx  = \int_{0}^{Q_{i}(p)}\{ 1 - F_{i}(x) \}dx,  
\end{equation}
uniformly in $p \in [0, 1]$. The quantity in the right side of Equation \eqref{eq:TTTlim} is termed as TTT and denoted by $T_{i}(p)$.

\subsection{Scalar-on-distribution function regression}

Assume $Y_{i}$ is the subject-specific scalar outcome of interests that could be continuous or discrete from an exponential family. We assume the following scalar-on-functional regression model with distributional representations such as distribution function, and quantile function as predictors. 
\begin{equation}
 \label{eq:mod}
 E\{ Y_{i}| X_{i_{1}}, \cdots, X_{im_{i}}\} = \mu_{i} \text{ where } g(\mu_{i}) = \alpha + \nu_{i} \text{ with }
 \nu_{i} = \int_{0}^{1} D_{i}(z)\beta(z)dz.
 \end{equation}
Here $g(\cdot)$ is a known link function and $D_{i}$ captures the distributional accumulation of $X_{i_{1}}, \cdots X_{i_{m_{i}}}$ as discussed in Section \ref{subsec:dist}. 
The smooth coefficient function $\beta(z)$ represents the functional effect of the transformation of the distribution function at level $z$. 
In the growing literature of functional data analysis, there are some available methods for the estimation and for the functional-coefficient in scalar-on-function, for example, see 
\cite{ramsay2005springer, reiss2007functional, goldsmith2011penalized} among many others. 
\par
In this article, we follow a smoothing spline estimation technique where we represent the unknown coefficient function $\beta(z)$ in terms of univariate basis functions expansion $\beta(z) = \sum_{k=1}^{\kappa}\theta_{k}B_{k}(z)$, where $\{B_{k}(z)\}_{k = 1}^{\kappa}$ is the set of known basis functions over $z$. 
In matrix notation, one can write $\beta(z) = \bB(z)^{\tp}\btheta$ where $\bB(z) = [B_{1}(z), \cdots, B_{\kappa}(z)]^{\tp}$ is the vector of known basis function and $\btheta = (\theta_{1}, \cdots, \theta_{\kappa})^{\tp}$ is the vector of unknown coefficients.
In this article, we have used B-spline due to its practicability \cite{de1978practical}. Therefore, the linear functional effect can be re-written as $\int D_{i}(z)\beta(z)dz = \sum_{k=1}^{\kappa}\theta_{k}\int D_{i}(z)B_{k}(z) dp= \sum_{k=1}^{\kappa}\theta_{k}W_{ik} = \bW^{\tp}\btheta$ where $W_{ik} = \{\int_{0}^{1}D_{i}(z)B_{k}(z)dz \}$ and $\bW_{i} = (\bW_{i1}, \cdots, \bW_{i\kappa})^{\tp}$.
\par
Thus, from the above discussion, we can rewrite Equation \eqref{eq:mod} as $g(\mu_{i}) = \alpha + \bW_{i}^{\tp}\btheta$ where the unknown is $\alpha, \btheta$, to estimate them, use the penalized log-likelihood criterion as following. 
\begin{equation}
    \sR(\alpha, \beta(\cdot)) = -2\log\sL(\alpha,  \btheta; Y_{i}, \bW_{i}) + \sP(\btheta)
\end{equation}
where $\sP$ is the penalty based on the smoothness of the coefficient function. 
We define $\sP = \lambda \int \{\beta''(z) \}^{2}dz = \lambda\btheta^{\tp}\bbP\btheta$ where $\beta''$ denotes the second derivative of the smooth function $\beta$ and the penalty matrix $\bbP$ is given by $\bbP = \int \btheta''(z)^{\tp}\btheta''(z)dz$. 
Therefore, we implement the minimization of $\sR$ using the Newton-Raphson method under the generalized additive models \cite{wood2016smoothing, wood2017generalized}. 
The smoothing parameter $\lambda$ is chosen using REML, information criteria such as AIC, BIC, or data-driven techniques such as generalized cross-validation. For implementation using popular statistical software R \cite{R}, we use \texttt{refund} package \cite{refund}. 

\begin{remark}[Rationale behind using TTT]
Numerous alternatives exist for the selection of $D_{i}$, encompassing options like DF or QF \cite{ghosal2021distributional}. 
Given such an array of possibilities, when $\beta(p) = \beta$ holds as a constant, the relationship $\nu_{i} = \beta\int_{0}^{1}D_{i}(p)dp = \beta\E\{X_{i}\}$ becomes apparent. This, in turn, leads to the simplification of the model described in  Equation \eqref{eq:mod} into the conventional generalized linear model with scalar predictors and responses. 
Consequently, the consideration of the distributional aggregation of $X_{ij}$s is overlooked, resulting in a compromise of predictive performance.
Moreover, due to the mathematical properties of total time on the test,
$T_{i}(1)$ is the same as the mean of the underlying distribution, viz., $\nu_{i} = \beta\times T_{i}(1)$. 
Rather than focusing solely on the endpoint of a function, a more insightful approach involves incorporating the complete function as a predictor within Model \eqref{eq:mod}. 
This can be achieved through a generalized form of $\nu_{i} = \int_{0}^{1}\beta(p)T_{i}(p)dp$, offering a broader perspective for prediction.
\end{remark}

\section{Application to HEAL-MS study}
\label{sec:data-analysis}
In this section, we present detailed analyses of the accelerometry data from the HEAL-MS study to compare the discriminatory performance of five distributional representations. We model two outcomes: (i) disability as a binary variable (low vs. high EDSS) and (ii) original EDSS scores. In both cases, we use all five distributional representations and scalar average activity counts within 8am-8pm time interval, obtained using both original and log-transformed activity counts. 

We estimated quantile and TTT functions over percentile levels, while density, survival, and hazard functions are approximated across the entire range of minute-level activity counts. Figures \ref{fig:AC} and \ref{fig:LAC} show the scalar and distributional summaries of activity counts on both original and log-transformed scales, respectively. Bold solid lines represent the simple point-wise averages of the distributional representations for different groups. It is important to note that some of these representations serve primarily as point-wise summaries and may not convey deeper statistical meaning and a proper distributional representation.

\begin{figure}
    \centering
    \includegraphics[width = \textwidth]{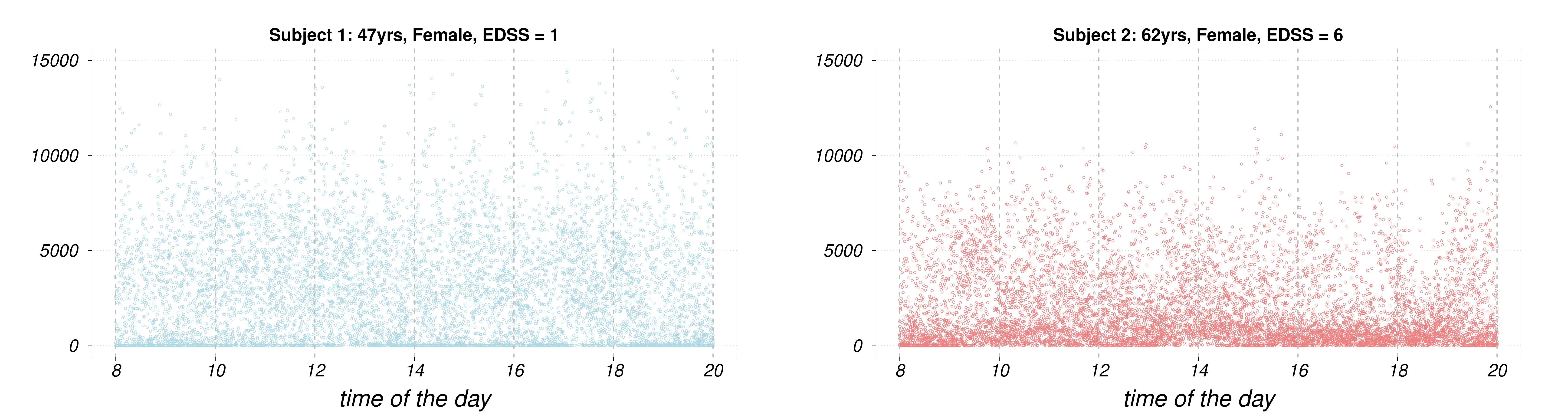}
    \includegraphics[width = \textwidth]{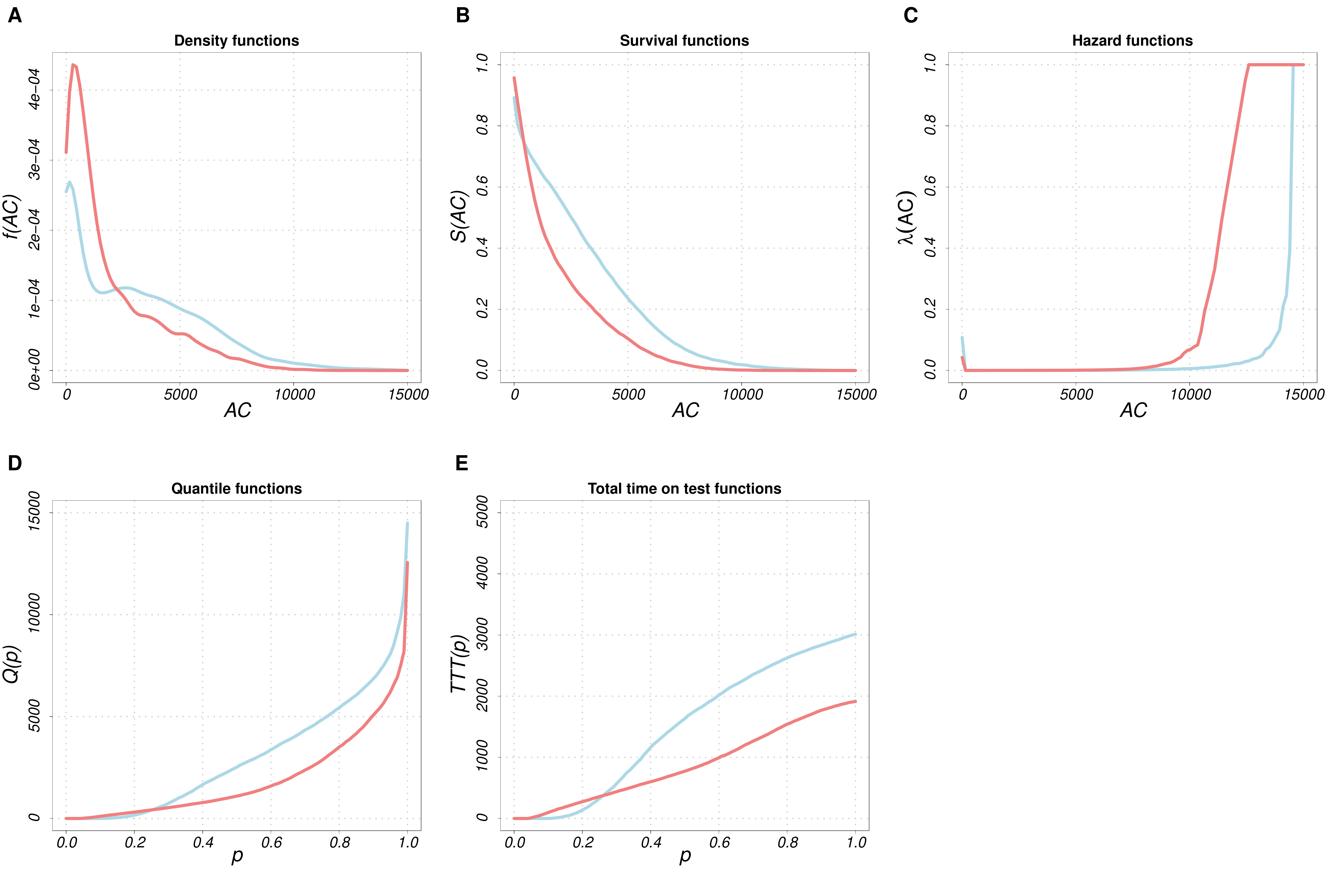}
    \caption{\underline{\textbf{Five distributional summaries of \textit{activity counts (AC)} for two subjects}:} Density functions (A), survival functions (B), hazard functions (C) with corresponding quantile functions (D), and corresponding TTT functions (E).
    }
    \label{fig:indiv-AC}
\end{figure}

\begin{figure}
    \centering
    \includegraphics[width = \textwidth]{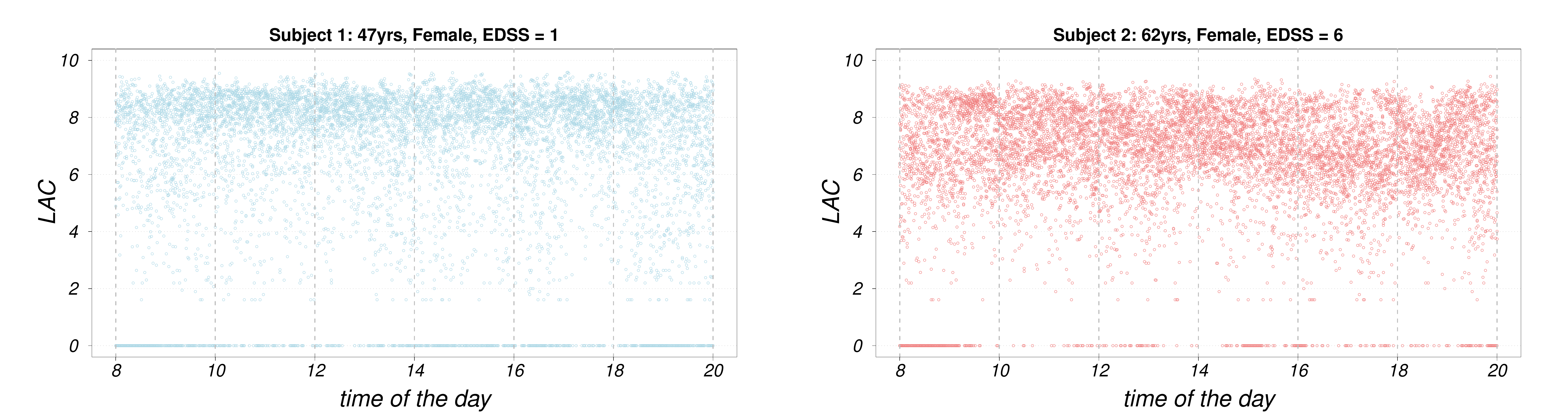}
    \includegraphics[width = \textwidth]{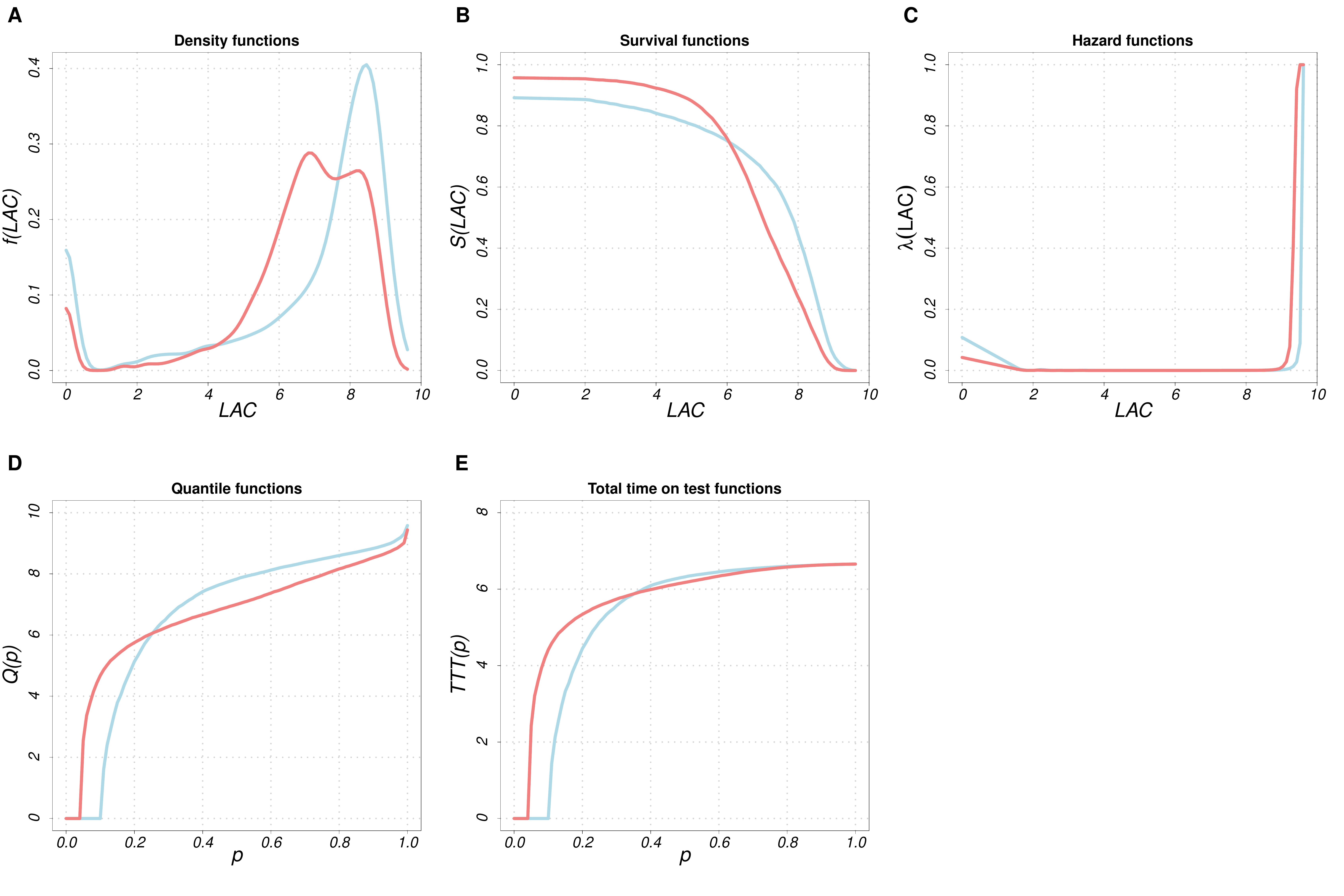}
    \caption{\underline{\textbf{Five distributional summaries of \textit{log-transformed activity counts (LAC)} for two subjects}:} Density functions (A), survival functions (B), hazard functions (C) with corresponding quantile functions (D), and corresponding TTT functions (E).
    }
    \label{fig:indiv-LAC}
\end{figure}

\begin{figure}
    \centering
    \includegraphics[width = \textwidth]{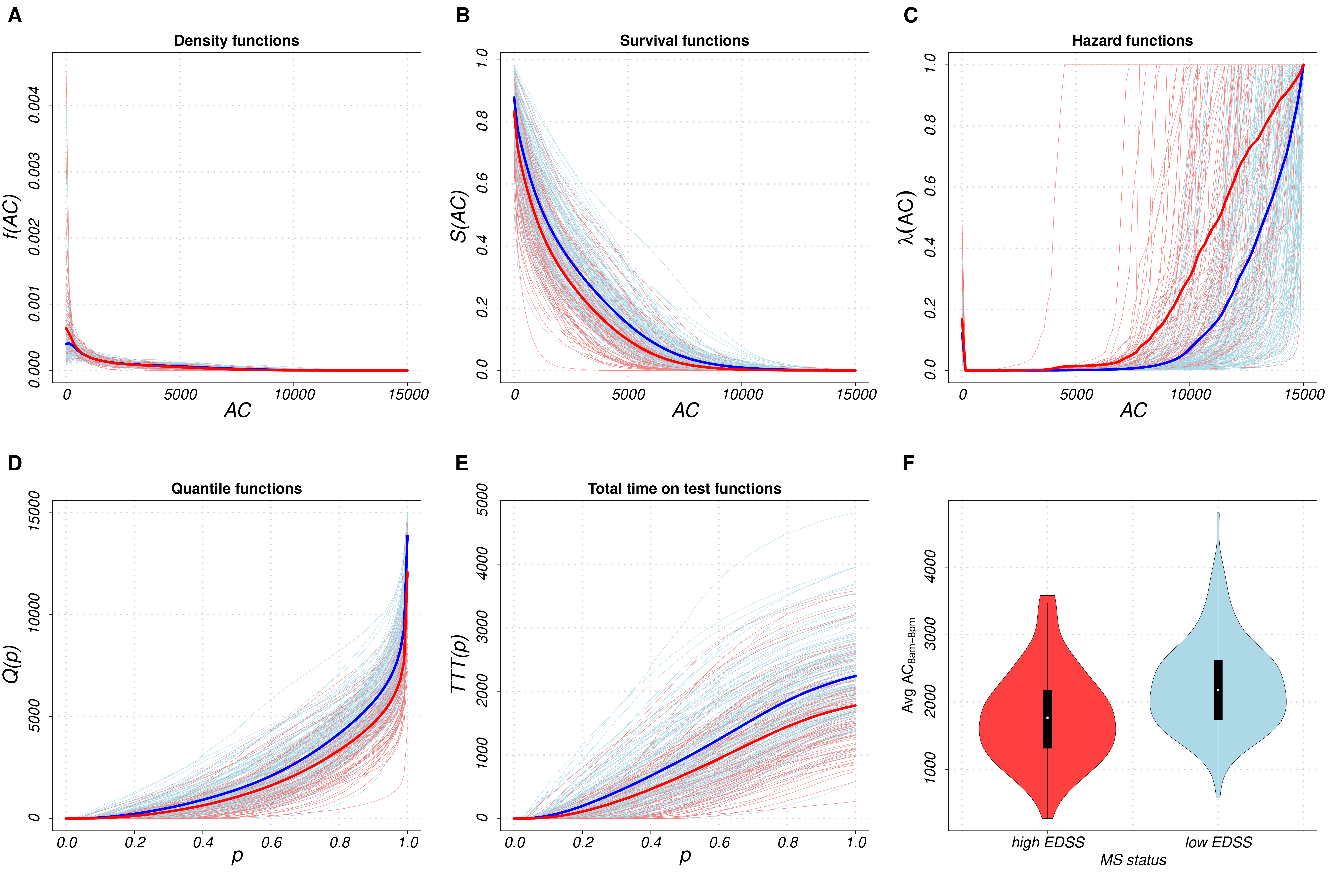}
    \caption{\underline{\textbf{ Distributional and scalar summaries of \textit{activity counts (AC)}:}} Density functions (A), survival functions (B), hazard functions (C) with corresponding quantile functions (D), and corresponding TTT functions (E). In addition, the violin-plot of scalar summaries is also shown in (F). 
    Two MS statuses based on EDSS (high vs low) are color-coded by light red and blue, respectively. The solid red and blue lines indicate the barycenter of the corresponding distributional representations for MS classes with low and high EDSS, respectively.}
    \label{fig:AC}
\end{figure}
\begin{figure}
    \centering
    \includegraphics[width = \textwidth]{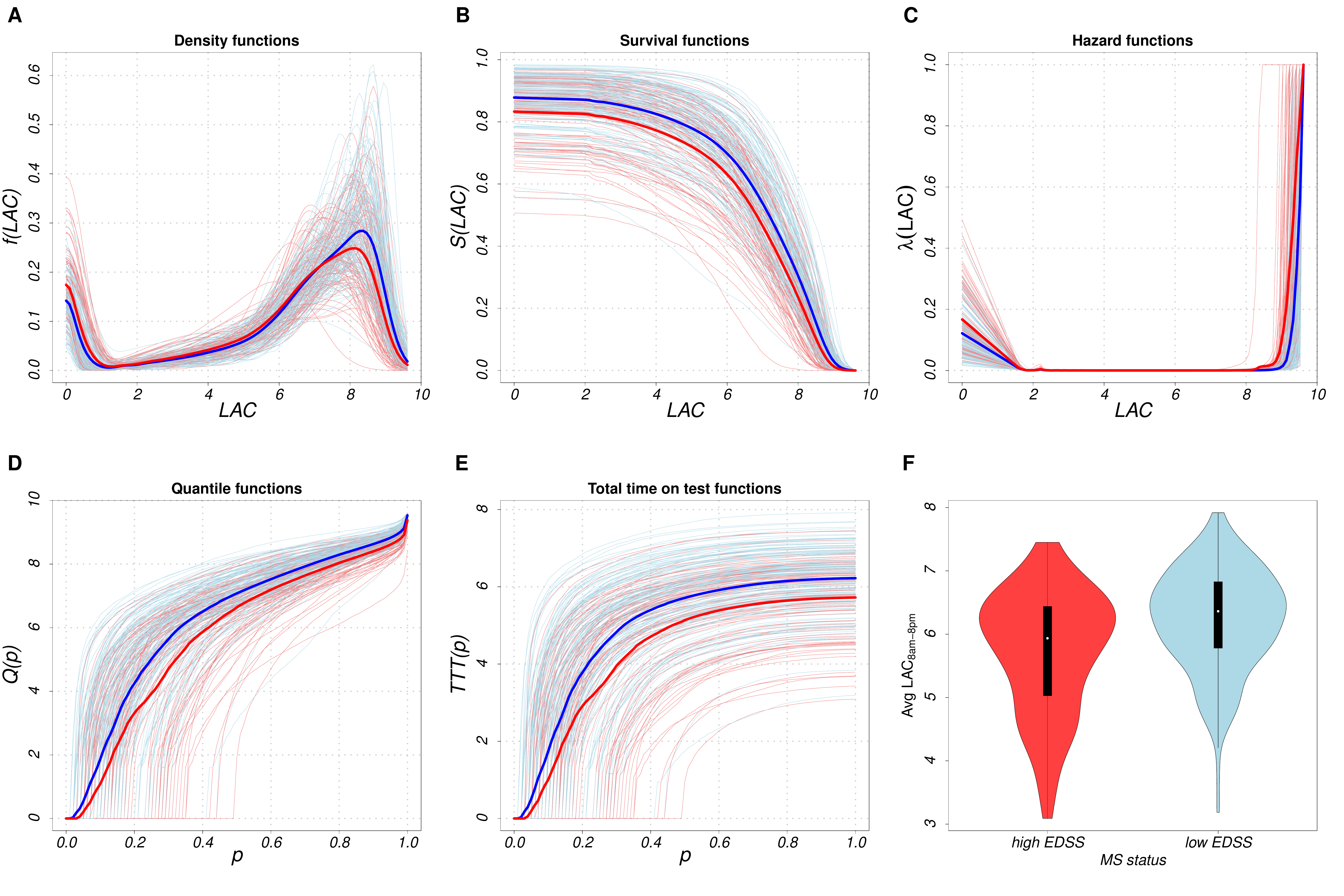}
    \caption{\underline{\textbf{Distributional and scalar summaries of \textit{log activity counts (LAC)}:}} Density functions (A), survival functions (B), hazard functions (C) with corresponding quantile functions (D), and corresponding TTT functions (E). In addition, the violin-plot of scalar summaries is also shown in (F). Two MS statuses based on EDSS (high vs low) are color-coded by light red and blue, respectively. The solid red and blue lines indicate the barycenter of the corresponding distributional representations for MS classes with low and high EDSS, respectively.}
    \label{fig:LAC}
\end{figure}
\par
To assess the discriminatory performance of the six predictors, we utilize cross-validated Area Under the Curve (cvAUC) for the binary and $cvR^2$ for the continuous outcome, respectively. To ensure the reliability of our validation, we employ a rigorous 5-fold cross-validation technique with 100 random replications and present the average cross-validated AUC and cross-validated $R^{2}$ along with corresponding 95\% confidence intervals to provide a comprehensive understanding of the model's performance in Table \ref{tab:cvAUC-MS}. 
\begin{table}
\centering
\caption{(i) Cross-validated AUCs for binary disability (high vs low EDSS) are shown in columns 2 and 3. (ii) Cross-validated $R^{2}$s for continuous disability (original EDSS scores) are shown in columns 4 and 5. The 95\% confidence intervals of are indicated in the parenthesis. All models are described by using the mean function, $\eta_{i}$ described in Section \ref{sec:model}.}
\label{tab:cvAUC-MS}
\begin{tabular}{ccccc}
\toprule
$\eta_{i}$ & \multicolumn{2}{c}{cross-validated AUC} & \multicolumn{2}{c}{cross-validated $R^{2}$} \\ 
\cmidrule{2-3}\cmidrule{4-5}
& AC in original scale & AC in log scale & AC in original scale & AC in log scale \\
\midrule
\multicolumn{5}{c}{Scalar summaries as covariates}\\
\hdashline
$\beta_{{AC}{[8am-8pm]}}\overline{\text{AC}}_{i[8am-8pm]}$ &
0.70 (0.69, 0.71) &
0.65 (0.64, 0.66) &
0.13 (0.11, 0.15) &
0.10 (0.08, 0.12)\\
\midrule
\multicolumn{5}{c}{Distribution summaries with domain in quantile levels as covariates}\\
\hdashline
$\int \beta_{T}(p)T_{i}(p)dp$ &
0.73 (0.72, 0.75) &
0.68 (0.64, 0.71) &
0.21 (0.19, 0.23) &
0.18 (0.13, 0.22)\\
$\int \beta_{Q}(p)Q_{i}(p)dp$ &
0.76 (0.74, 0.77) &
0.74 (0.72, 0.76) &
0.23 (0.22, 0.25) &
0.23 (0.21, 0.26)\\
\midrule
\multicolumn{5}{c}{Distribution summaries with domain in sample space as covariates}\\
\hdashline
$\int \beta_{f}(x)f_{i}(x)dx$ &
0.71 (0.69, 0.73) &
0.67 (0.64, 0.69) &
0.12 (0.10, 0.15) &
0.11 (0.08, 0.14)\\
$\int \beta_{S}(x)S_{i}(x)dx$ &
0.72 (0.70, 0.74) &
0.69 (0.67, 0.71) &
0.12 (0.10, 0.15) &
0.12 (0.09, 0.14)\\
$\int \beta_{\lambda}(x)\lambda_{i}(x)dx$ &
0.77 (0.76, 0.78) &
0.77 (0.76, 0.78) &
0.26 (0.23, 0.28) &
0.25 (0.24, 0.27)\\
\bottomrule
\end{tabular}%
\end{table}
\par
\begin{figure}
    \centering
    \includegraphics[width = \textwidth]{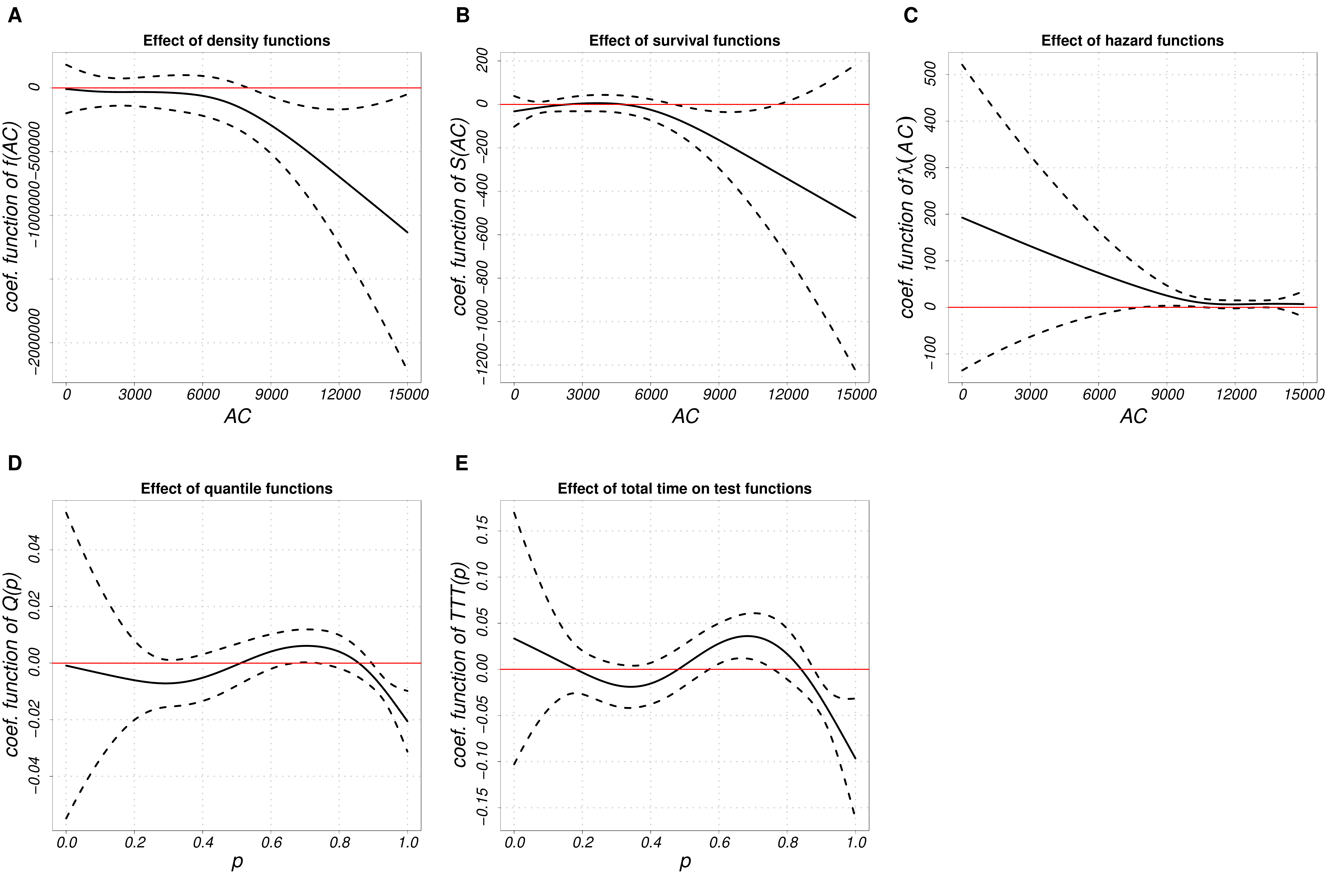}
    \caption{
    The estimated regression coefficients for the models with different distributional representations were obtained from scalar-on-function \textit{logistic regressions to model MS status}, where the distributional and scalar summaries are obtained based on \textit{activity counts}. A solid black curve indicates the estimated coefficient function, dotted black lines indicate 95\% confidence intervals.}
    \label{fig:coef-logistic-AC}
\end{figure}
\begin{figure}
    \centering
    \includegraphics[width = \textwidth]{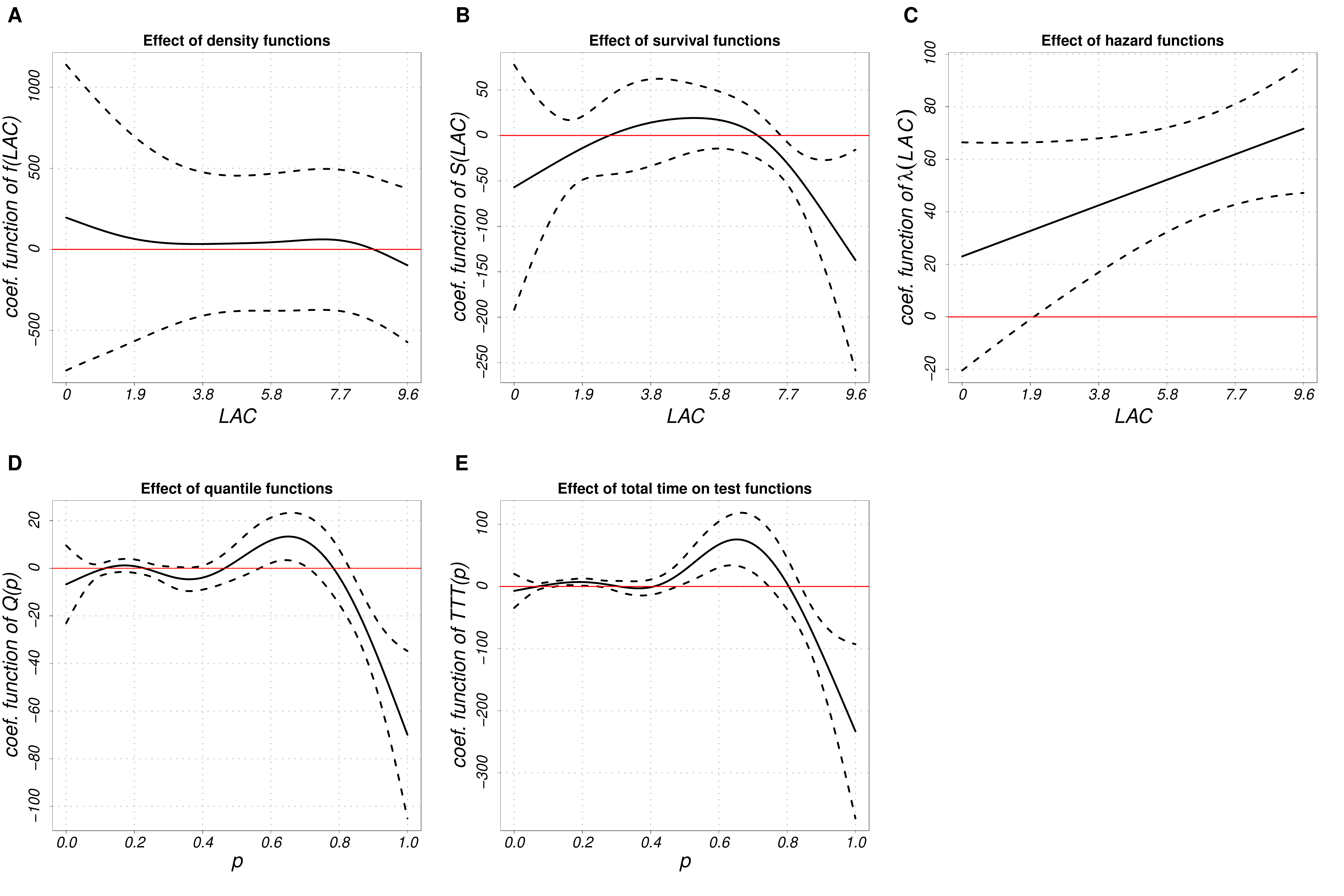}
    \caption{The estimated regression coefficients for the models with different distributional representations were obtained from \textit{scalar-on-function logistic regressions to model MS status}, where the distributional and scalar summaries are obtained based on \textit{log-activity counts}. A solid black curve indicates the estimated coefficient function, dotted black lines indicate 95\% confidence intervals.}
    \label{fig:coef-logistic-LAC}
\end{figure}
\begin{figure}
    \centering
    \includegraphics[width = \textwidth]{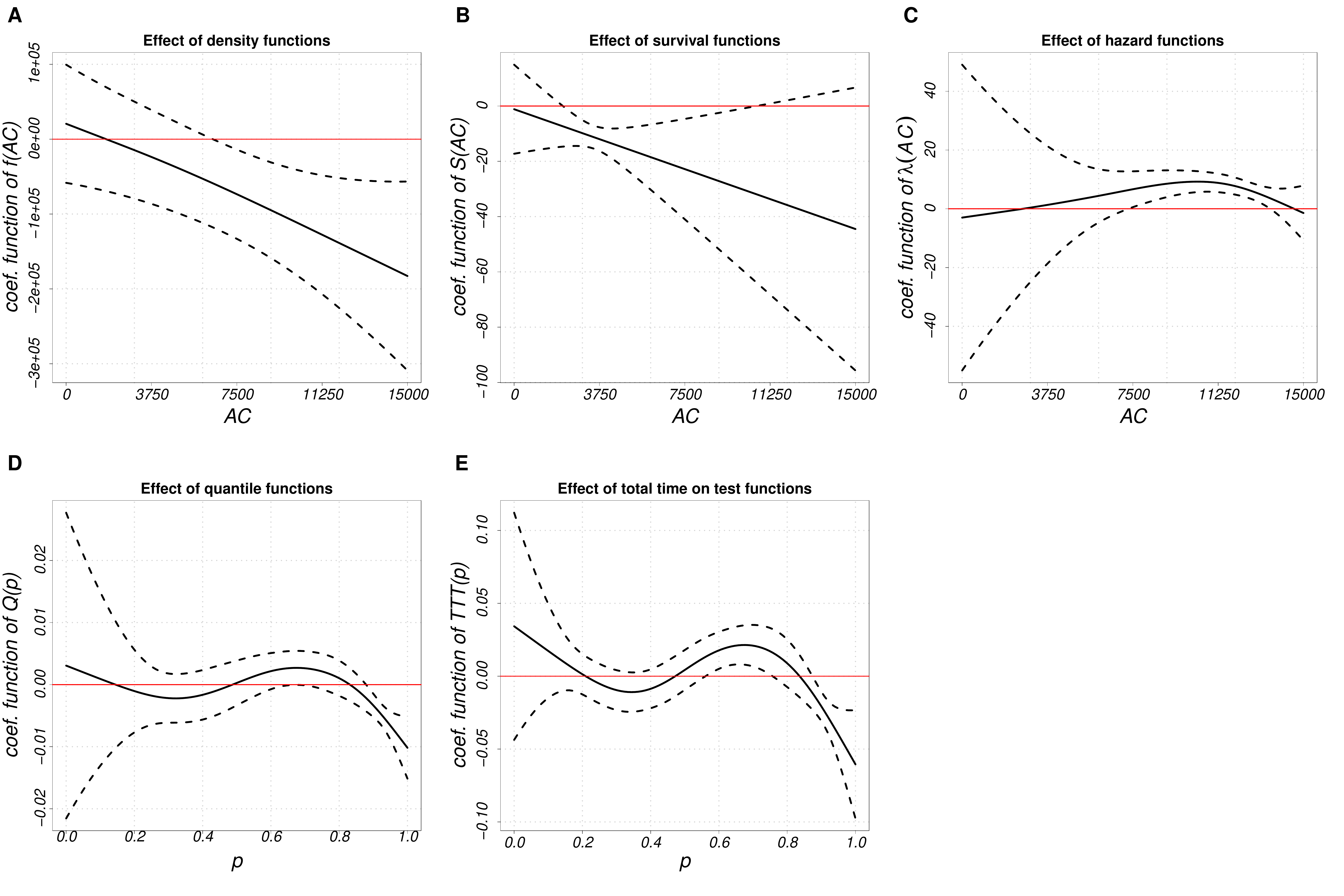}
    \caption{The estimated regression coefficients for the models with different distributional representations were obtained from \textit{scalar-on-function regressions to model EDSS scores}, where the distributional and scalar summaries are obtained based on \textit{activity counts}. A solid black curve indicates the estimated coefficient function, dotted black lines indicate 95\% confidence intervals. }
    \label{fig:coef-cont-edss-AC}
\end{figure}
\begin{figure}
    \centering
    \includegraphics[width = \textwidth]{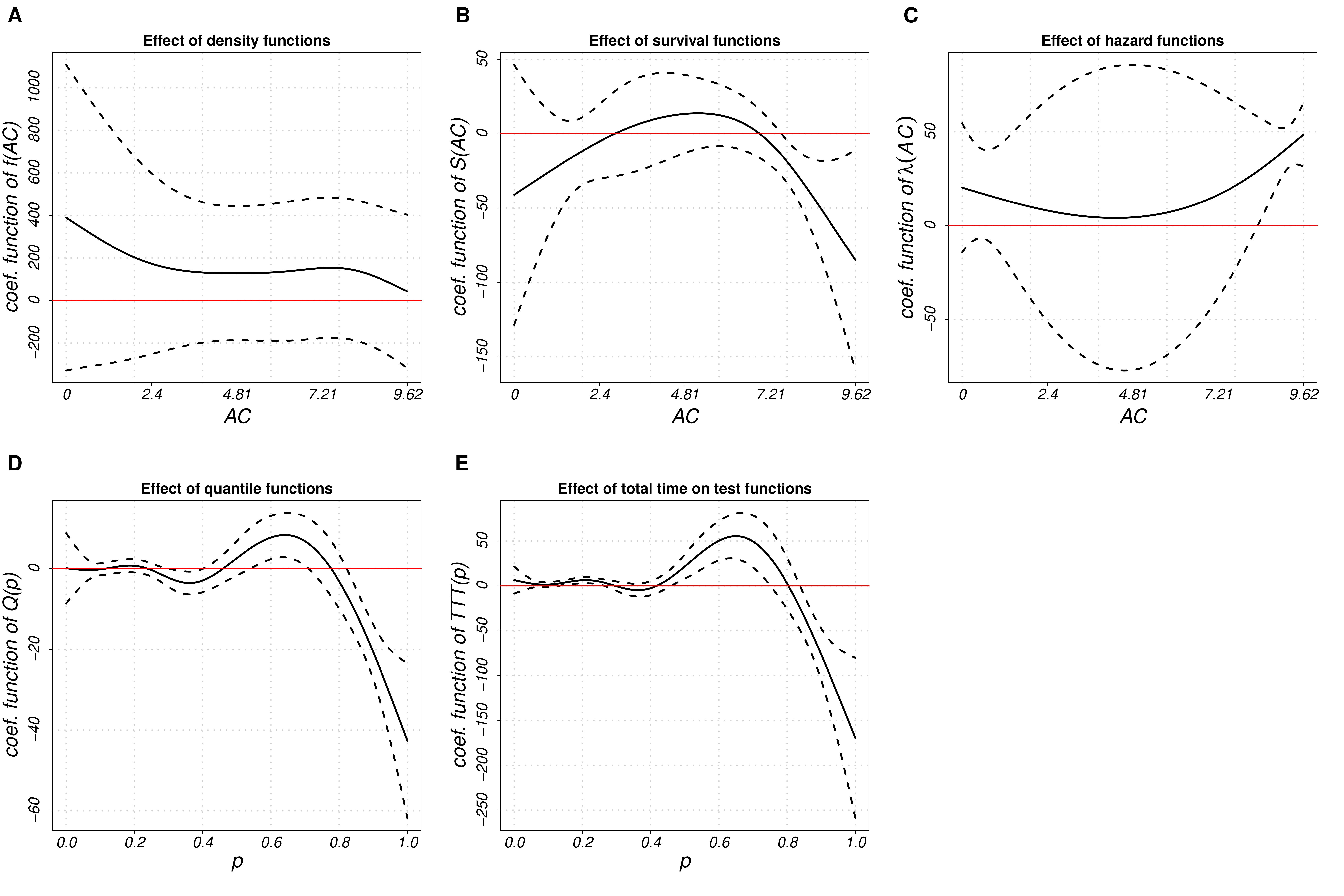}
    \caption{The estimated regression coefficients for the models with different distributional representations were obtained from \textit{scalar-on-function regressions to model EDSS scores}, where the distributional and scalar summaries are obtained based on \textit{log-activity counts}. A solid black curve indicates the estimated coefficient function, dotted black lines indicate 95\% confidence intervals.}
    \label{fig:coef-cont-edss-LAC}
\end{figure}

Figures \ref{fig:coef-logistic-AC} and \ref{fig:coef-logistic-LAC} show estimated regression coefficient functions obtained from the distributional scalar-on-function logistic regression models. Similarly, Figures \ref{fig:coef-cont-edss-AC} and \ref{fig:coef-cont-edss-LAC} show estimated regression coefficient functions obtained from the distributional scalar-on-function Gaussian regression models for predicting the EDSS where all distributional representations are calculated based on the original and log-transformed activity counts, respectively. 

Key findings are summarized as follows:
\begin{itemize}

\item[] {\bf Binary disability (High vs. Low EDSS):}

\item[*] The hazard function demonstrated the highest discriminatory power, with a cvAUC of 0.77 (95\% CI: 0.76–0.78) when using original activity counts and 0.77 (95\% CI: 0.76–0.78) with log-transformed counts.

\item[*] The quantile function provided the second-highest performance, achieving a cvAUC of 0.76 (95\% CI: 0.74–0.77).
Conventional scalar summaries showed significantly lower performance, with cvAUCs of 0.70 (95\% CI: 0.69–0.71) and 0.65 (95\% CI: 0.64–0.66) for original and log-transformed counts, respectively.
\item[]
\item[] { \bf Continuous disability (original EDSS scores):}

\item[*] The hazard function again demonstrated superior performance, achieving an $cvR^2$ of 0.26 (95\% CI: 0.23–0.28) for original counts and 0.25 (95\% CI: 0.24–0.27) for log-transformed counts.
\item[*] The quantile and TTT functions showed comparable performance, with $cvR^2$ values ranging in 0.23–0.24.
Scalar summaries performed poorly, with $cvR^2$ values of 0.13 (original AC) and 0.1 (log-transformed AC).
\end{itemize}

Additional observations are as follows.
\begin{itemize}
    \item Modeling MS status based on activity counts (AC): From the output of the classical logistic regression, we find that the subject-specific average PA is significantly associated (with the level of significance $\alpha = 0.05$) with lower log odds of MS, 
    $\widehat{\beta}_{AC_{[8am-8pm]}} = -0.001~(0.0002)$. 
    Figure \ref{fig:coef-logistic-AC} further illustrates the distribution effects of log-odds of MS based on AC.  
 
    \item Modeling MS status based on log-activity counts (LAC): Using the classical logistic regression, we find the subject-specific average PA is significantly associated with lower log odds of MS $\widehat{\beta}_{AC_{[8am-8pm]}} = -0.65~(0.11)$. Figure \ref{fig:coef-logistic-LAC} illustrates the distribution effects of log-odds of MS based on LAC.
    Statistically, both the quantile and TTT function are significantly associated with lower odds of MS, survival of higher LAC is significantly associated with lower odds of MS, and survival of higher LAC is significantly associated with higher odds of MS. 
    
    \item Modeling EDSS score based on AC:  Based on the classical regression model, we obtain subject-specific average PA is significantly associated with EDSS $\widehat{\beta}_{AC_{[8am-8pm]}} = -0.001~(0.0001)$. In figure \ref{fig:coef-cont-edss-AC}, we demonstrate the distributional effect of EDSS based on AC. The effect of the density function is significantly associated with lower EDSS, hazard for lower PA is significantly associated with lower EDSS. Higher levels of quantile and TTT functions are associated with lower EDSS. 
    
    \item Modeling EDSS score based on LAC: Based on the classical regression model, we obtain subject-specific average PA is significantly associated with EDSS $\widehat{\beta}_{AC_{[8am-8pm]}} = -0.51 ~(0.11)$. In figure \ref{fig:coef-cont-edss-LAC}, we demonstrate the distributional effect of EDSS based on log-transformed AC. The effect of survival for higher PA is significantly associated with lower EDSS. Similar to the previous situation, higher levels of quantile and TTT functions are associated with lower EDSS.
\end{itemize}

\par
\begin{figure}
    \centering
    \includegraphics[width = \textwidth]{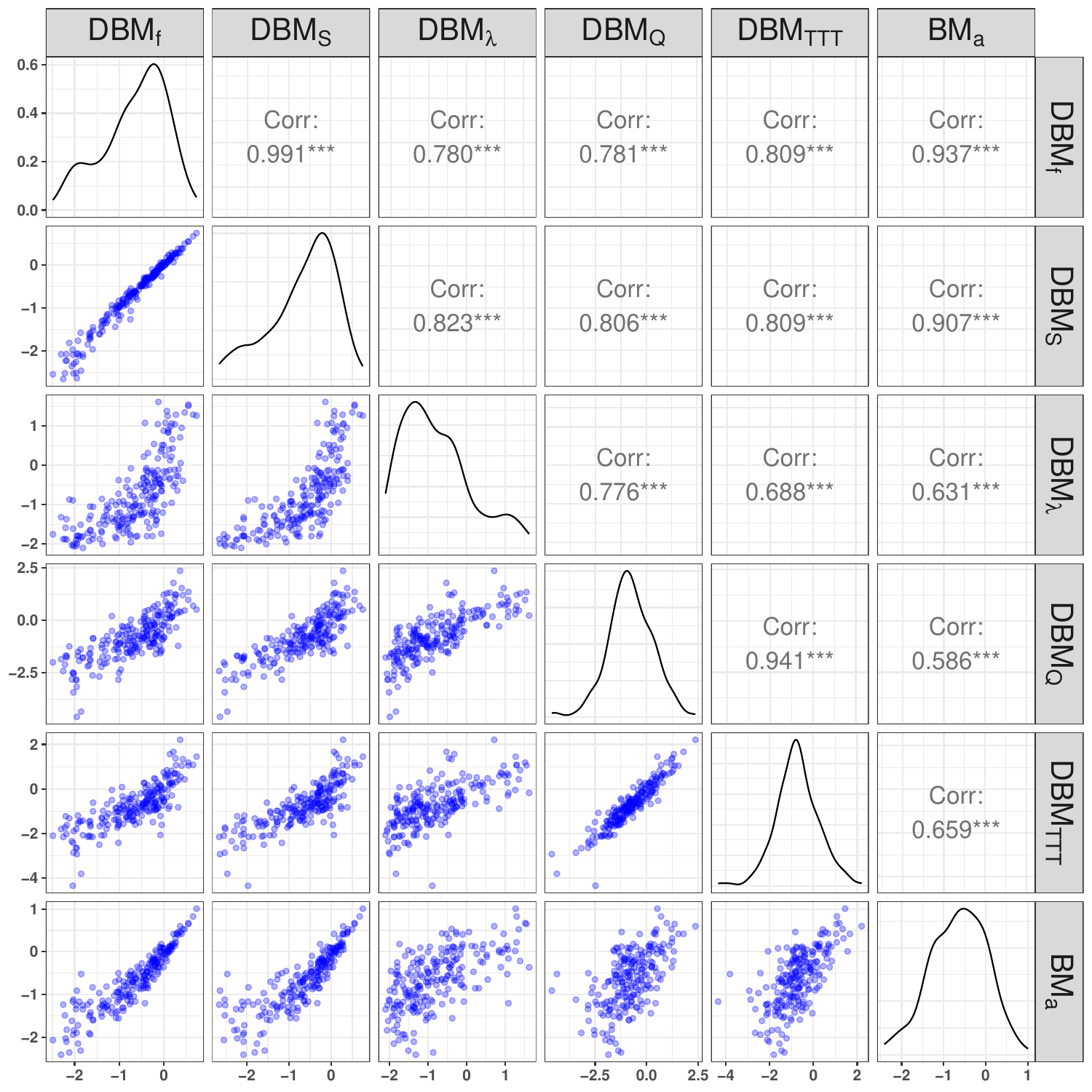}
    \caption{
    Scatter plots of the estimated weighted scores corresponding to the covariates density, survival, hazard, quantile, and TTT functions along with the average physical activity counts to \textit{predict the MS status}, where the distributional and scalar summaries are obtained based on \textit{activity counts}. The upper triangular entries indicate the Spearman's correlation coefficients between the associated row and columns.}
    \label{fig:bm-edss}
\end{figure}
\begin{figure}
    \centering
    \includegraphics[width = \textwidth]{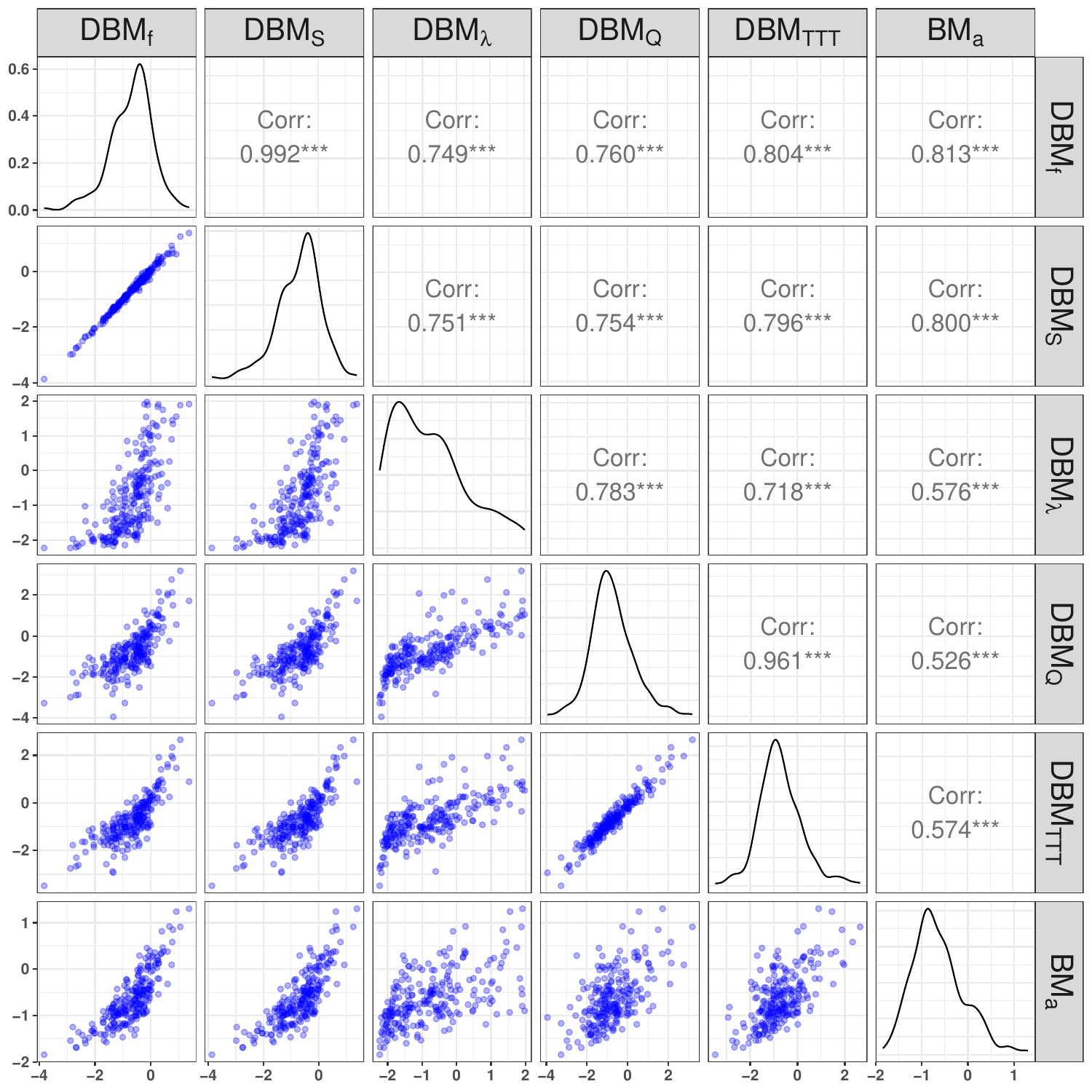}
    \caption{ 
    Scatter plots of the estimated weighted scores corresponding to the covariates density, survival, hazard, quantile, and TTT functions along with the average physical activity counts to \textit{predict the EDSS scores}, where the distributional and scalar summaries are obtained based on \textit{log-activity counts}.
    The upper triangular entries indicate the Spearman's correlation coefficients between the associated row and columns.
    }
    \label{fig:bm-edss-log}
\end{figure}
\begin{figure}
    \centering
    \includegraphics[width = \textwidth]{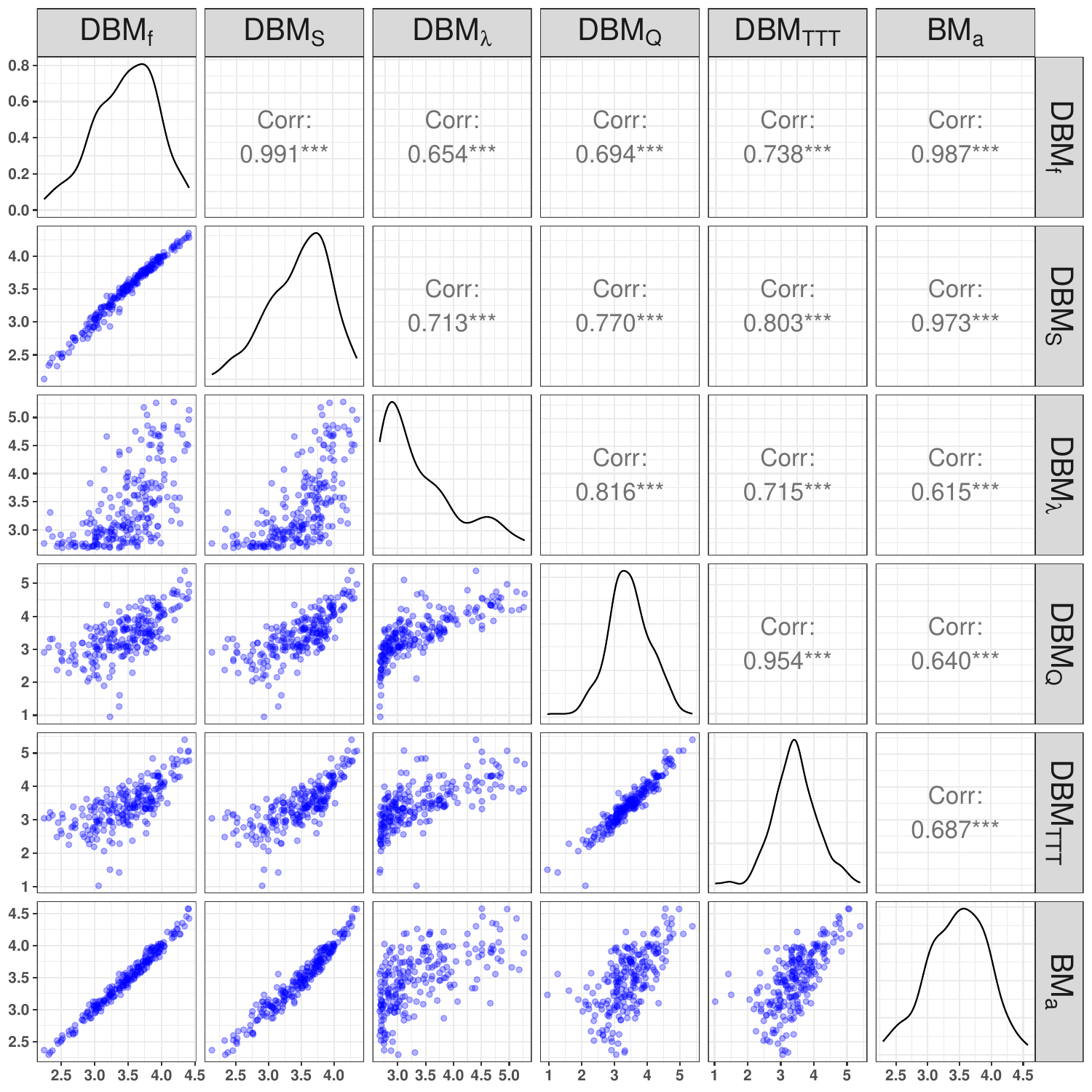}
    \caption{
    Scatter plots of the estimated weighted scores corresponding to the covariates density, survival, hazard, quantile, and TTT functions along with the average physical activity counts to \textit{predict EDSS scores}, where the distributional and scalar summaries are obtained based on \textit{activity counts}. The upper triangular entries indicate the Spearman's correlation coefficients between the associated row and columns.}
    \label{fig:bm-cont-edss}
\end{figure}
\begin{figure}
    \centering
    \includegraphics[width = \textwidth]{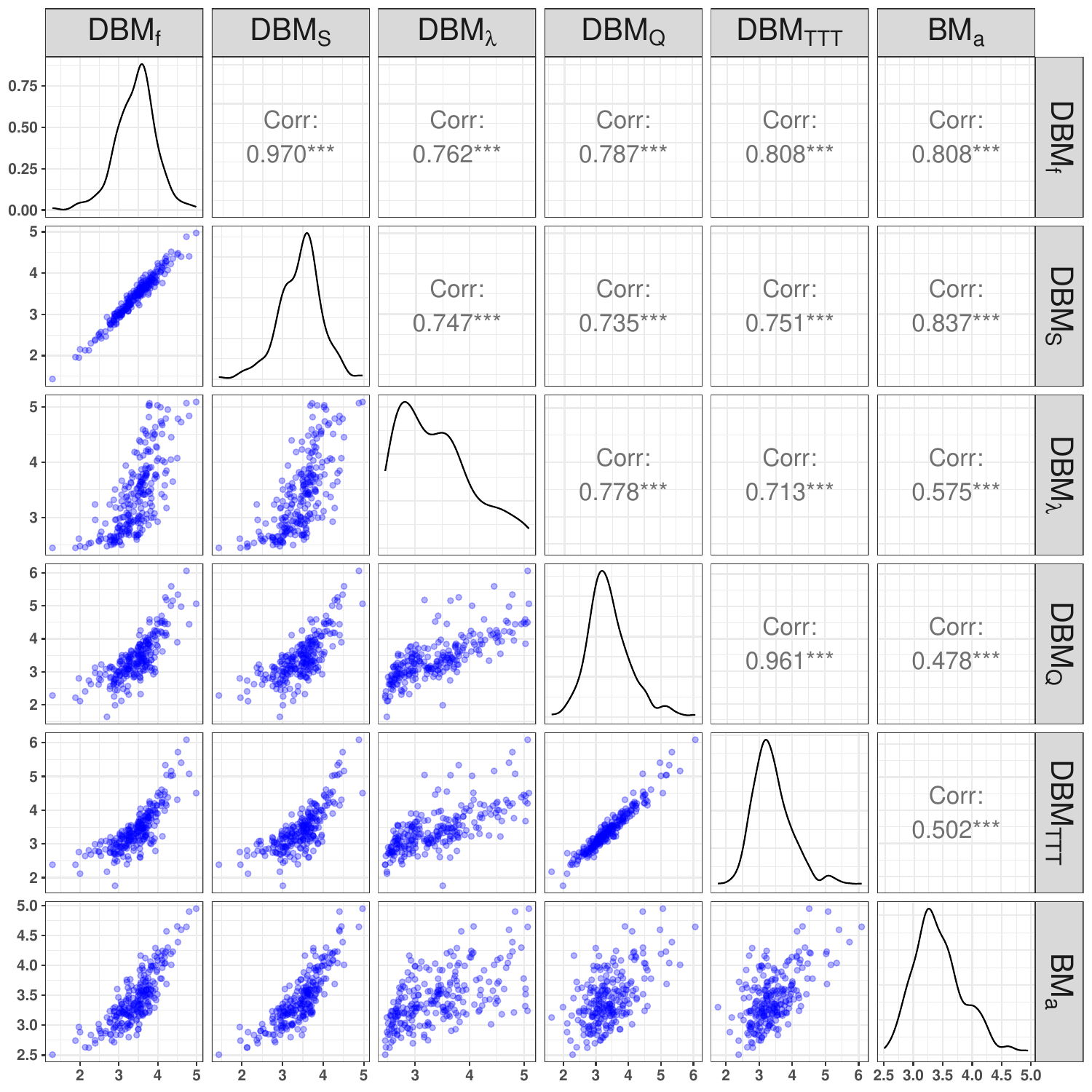}
    \caption{
    Scatter plots of the estimated weighted scores corresponding to the covariates density, survival, hazard, quantile, and TTT functions along with the average physical activity counts to predict EDSS, where the distributional and scalar summaries are obtained based on \textit{log-activity counts}. The upper triangular entries indicate the Spearman's correlation coefficients between the associated row and columns.}
    \label{fig:bm-cont-edss-log}
\end{figure}

Based on the outcome of the fitted model, we are also interested in calculating the regression biomarkers and comparing them across distributional representations. The biomarkers based on the models with density, survival, hazard, quantile, TTT functions, and average physical activity as covariates are defined as follows. 
(i) $DBM_{f, i} = \int f_{i}(x)\widehat{\beta}_{f}(x)dx$,
(2) $DBM_{S, i} = \int S_{i}(x)\widehat{\beta}_{S}(x)dx$,
(3) $DBM_{\lambda, i} = \int \lambda_{i}(x)\widehat{\beta}_{\lambda}(x)dx$, 
(4) $DBM_{Q, i} = \int Q_{i}(p)\widehat{\beta}_{Q}(p)dp$,
(5) $DBM_{TTT, i} = \int T_{i}(p)\widehat{\beta}_{T}(x)dp$ and (6) $BM_{a, i} = \overline{\text{AC}}_{i}\widehat{\beta}_{AC_{i,[8am-8pm]}}$. Figures \ref{fig:bm-edss}, \ref{fig:bm-edss-log} demonstrate the scatter-plot matrices for all five types of biomarkers to discriminate the MS status where the distributional and scalar summaries are obtained from counts in original and log scale, respectively. In addition, similar to the earlier case, Figures \ref{fig:bm-cont-edss} and \ref{fig:bm-cont-edss-log} display the scatter-plot matrices for all five biomarkers to discriminate the EDSS where the distributional and scalar summaries are obtained from counts in original and log scale, respectively. We can clearly see that the biomarkers fall into three distinct groups based on their cross-correlations. Density and CDF-based biomarkers are very similar between each other and are highly correlated with the average activity counts. On the other hand, quantile and TTT-based biomarkers are highly correlated, capturing very similar information across quantile levels. {\em Hazard-based biomarkers stand out by uniquely capturing tail information not captured by the other four unconditional distributional representations.}

\section{Discussion}
\label{sec:dis}
In this paper, we compared five individual-specific distributional representations to capture distributional information in the physical activity assessed by accelerometry. We developed scalar-on-function regression models using the proposed distributional representations, including unconditional probability-based domains and quantile domains. We demonstrate the benefits of modeling the distributional aspect of accelerometry data that significantly increased the power to discriminate disability levels in individuals with MS. Among the distributional representations explored in the HEAL-MS study, the hazard function demonstrated the highest discriminatory power.

From a clinical perspective, more sensitive distributional representation aggregating daily activity can help accelerate the conduct of clinical trials of candidate neuroprotective medications. This is important since the approved medications for MS do not appear to work in the progressive phase of the illness in the absence of ongoing attacks on myelin. 

From a mathematical perspective, all five distributional representations could be re-expressed in terms of each other. However, from a predictive modeling perspective, we are often interested in using linear functional models. This may favor certain distributional representations, as illustrated in our exploratory case study.

This research highlights the value of distributional information as a novel highly sensitive biomarker and endpoint in clinical studies, offering a more accurate way to capture within-subject variability in accelerometry data. Our group will apply these endpoints to track the longitudinal change in disability of the participants of the HEAL-MS study. Similar approaches can be used in data generated by other sDHT modalities including continuous glucose monitoring and heart rate monitors. In future work, we plan to incorporate temporal and longitudinal information in the characterization of distributions.

\section*{Acknowledgements}
This research was partially supported by the grant R01NR018851 from the National Institute of Health (NIH). 

\section*{Financial disclosure}
EM disclosed research support from Biogen, Roche/Genentech. Consulting fees from BeCareLink, LLC. Royalties for editorial duties from UpToDate.

\section*{Conflict of interests}
The authors declare no potential conflicts of interests

\bibliography{paper-main}

\end{document}